\documentclass[a4paper,english,aps,prl,twocolumn,showpacs,superscriptaddress,groupedaddress]{revtex4}
\usepackage[T1]{fontenc}
\usepackage[latin9]{inputenc}
\usepackage{babel}
\usepackage{amstext}
\usepackage{amssymb}
\usepackage{graphicx}
\usepackage[unicode=true,pdfusetitle,
 bookmarks=true,bookmarksnumbered=false,bookmarksopen=false,
 breaklinks=false,pdfborder={0 0 1},backref=false,colorlinks=false]
 {hyperref}

\makeatletter

\pdfpageheight\paperheight
\pdfpagewidth\paperwidth

\@ifundefined{textcolor}{}
{%
 \definecolor{BLACK}{gray}{0}
 \definecolor{WHITE}{gray}{1}
 \definecolor{RED}{rgb}{1,0,0}
 \definecolor{GREEN}{rgb}{0,1,0}
 \definecolor{BLUE}{rgb}{0,0,1}
 \definecolor{CYAN}{cmyk}{1,0,0,0}
 \definecolor{MAGENTA}{cmyk}{0,1,0,0}
 \definecolor{YELLOW}{cmyk}{0,0,1,0}
 }

\makeatother

\begin{document}

\title{}

\title{Circular dichroism of cholesteric polymers and the orbital angular
momentum of light}

\author{W. Löffler}

\email{loeffler@physics.leidenuniv.nl}

\affiliation{Huygens Laboratory, Leiden University, P.O. Box 9504, 2300 RA Leiden,
The Netherlands}

\author{D. J. Broer}

\affiliation{Eindhoven University of Technology, Dept. Functional Organic Materials
and Devices, P.O. Box 513, 5600 MB Eindhoven, Netherlands}

\author{J. P. Woerdman}

\affiliation{Huygens Laboratory, Leiden University, P.O. Box 9504, 2300 RA Leiden,
The Netherlands}
\begin{abstract}
We explore experimentally if the light's orbital angular momentum
(OAM) interacts with chiral nematic polymer films. Specifically, we
measure the circular dichroism of such a material using light beams
with different OAM. We investigate the case of strongly focussed,
non-paraxial light beams, where the spatial and polarization degrees
of freedom are coupled. Within the experimental accuracy, we cannot
find any influence of the OAM on the circular dichroism of the cholesteric
polymer.
\end{abstract}
\maketitle
Molecular chirality is of very high importance in biology, chemistry,
and material science. Materials with molecular chirality can be investigated
by optical means, because chiral molecules exhibit optical activity:
Their interaction with light is sensitive to the circular polarization
or helicity of the photons. This interaction is enantiomerically specific,
thus giving information about the structure of chiral matter (see,
e.g., \cite{barron2004}). One manifestation of optical activity is
circular dichroism, i.e., light absorption sensitive to the handedness
of the circular polarization. Optical activity is thus intrinsically
linked to the \emph{spin} angular momentum (SAM) of the photons. More
recently, it was recognized that photons can additionally carry \emph{orbital}
angular momentum (OAM). Such photons are naturally appearing in Laguerre-Gaussian
laser modes, where each photon carries $\ell\hbar$ of OAM ($\ell$
is the azimuthal mode index \cite{allen1992}). Does this additional
degree of freedom play a role in circular dichroism? 

If such interaction would be found, it is potentially useful for a
broad range of research areas and applications, comparable to (spin-based)
optical activity. Along this line,  the electromagnetic interaction
of OAM photons with atoms and molecules has received considerable
attention. However, the situation is unclear: The interaction of OAM
light with atoms and molecules has been studied in a number of theoretical
papers, with controversial outcomes: Some predict that such an interaction
should be observable within the electric dipole approximation \cite{alexandrescu2006,jauregui2004},
and some do not find such effects \cite{romero2002,andrews2003,andrews2004,babiker2002}.
The latter outcome is supported by the only experimental investigation
by Araoka et al. \cite{araoka2005}. In this experiment, the authors
tested if the molecular circular dichroism (CD) of an optically active
sample would be modified by using OAM probe light, with negative results.
Apparently, in a molecular system, OAM does not participate in the
same way in optical activity as SAM does. 

\begin{figure}
\includegraphics[width=1\columnwidth]{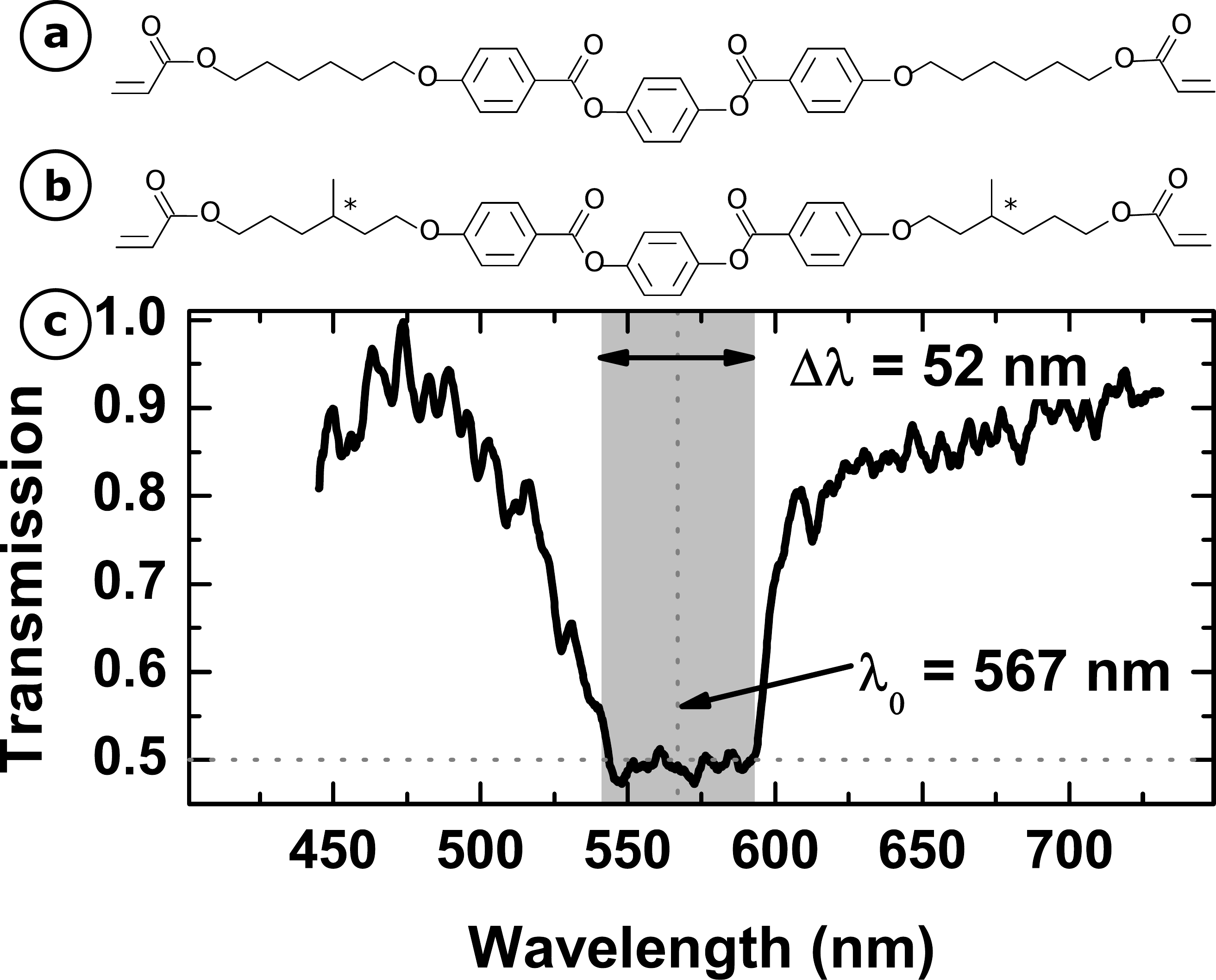}

\caption{\label{fig:clc}Structural formulas of the mesogens that we used (a,b).
Unpolarized transmission spectrum of the cholesteric polymer film
(c). In the reflection band (gray box) the wavelength matches the
pitch of the sample, there the reflected light's polarization has
the same handedness as the cholesteric polymer, whereas the opposite
polarization is transmitted.}
\end{figure}
Since OAM is a spatial property, it makes sense to study if OAM has
an effect in a system where the main contribution to optical activity
does not stem from molecular chirality but from spatial resonances.
Specifically, we address the structural chirality of a chiral nematic
polymer, a polymerized variant of a cholesteric liquid crystal. In
short-pitch chiral films, the alignment of the director has the same
symmetry (helical) and periodicity (the optical wavelength) as the
electric field vector in circularly polarized light (or as the wavefronts
in OAM light). This results in (spin-based) optical activity effects
orders of magnitude larger than in molecules. In this paper we investigate
experimentally if OAM influences the circular dichroism of such a
material. 

 We use a highly non-paraxial light beam so that the total angular
momentum depends on the SAM and OAM in a non-separable way \cite{barnett1994},
this coupling emerges from spin-orbit interaction of light \cite{bliokh2010}.
The coupling strength is proportional to $\theta_{0}^{2}/4$, where
$\theta_{0}$ is the (half-) aperture angle of the focussed beam \cite{nieminen2008}.
We can express the total angular momentum flux per unit length $J_{z}$,
normalized to the energy flux per unit length $E$ as: 
\[
\frac{J_{z}}{E}=\frac{\sigma_{z}+\ell}{\omega}+\frac{\sigma_{z}}{\omega}\left(\frac{4/\theta_{0}^{2}}{2p+\ell+1}+1\right)^{-1}
\]

$\sigma_{z}$ is the spin ($\pm1$), $\omega$ the angular frequency,
and $\ell$ and $p$ the azimuthal and radial indices of the Laguerre-Gaussian
mode. For small NA beams, the second term is negligible as in the
case of Araoka \cite{araoka2005}. We use a beam with $NA=0.55$,
this results in $\theta_{0}^{2}/4=0.1$, i.e., contributions to the
total angular momentum from terms involving both SAM and OAM are significant
\footnote{In the experiment of Araoka et al. \cite{araoka2005} the NA of the
beam was apparently much smaller than in our case.%
}.

Our cholesteric polymer films are based upon chiral nematic mixtures
\cite{broer1990,hikmet1991,lub1995} of the nematic reactive mesogen
1 and a chiral nematic reactive mesogen 2 (Fig.~\ref{fig:clc}a,
b). Mesogens are monomers, which, if polymerized, show similar properties
as liquid crystals. In our mixture, the wavelength of reflection can
be adjusted by their mixing ratio. The materials are mixed in a 1:1
weight ratio to give reflection in the green part of the spectrum
(Fig.~\ref{fig:clc}c). For photopolymerization, the isotropic photoinitiator
Irgacure 369 (IRG369, Ciba) was added in a quantity of 1~w\%. To
align the molecular layers at the bottom and top, a rubbed polyimide
coating was used; to average out retardation effects, the rubbing
directions were perpendicular with respect to each other. The thickness
of the film was controlled by using 16 $\mathrm{\mu m}$ spacers.
The polymer was filled in the cell by using capillary forces and polymerized
in its chiral nematic phase (at 85~$\mathrm{^{\circ}C}$). Such a
cholesteric polymer consists of helically oriented planes of parallel
aligned molecules, and the pitch $p_{0}$ corresponds to a full 2$\pi$
rotation of the director. This results in a Bragg-type reflection
band which is sensitive to the circular polarization \cite{stjohn1995,berreman1970}:
The polarization component with the same handedness as the director
helix is reflected, while the other component interacts only weakly
with the polymer and is transmitted. The polarization-averaged transmission
spectrum (Fig.~1) shows therefore 50\% transmission in the reflection
band. From a nonchiral variant of the polymer, the ordinary and extraordinary
refractive indices have been determined to be $n_{o}\approx1.55$
and $n_{e}\approx1.70$, respectively (at $\lambda=500\, nm$). From
the full width of this band $\Delta\lambda=p_{0}\Delta n$, where
$\Delta n=|n_{e}-n_{o}|$, we can determine the pitch $p_{0}=347$~nm,
this agrees very well with the pitch determined from the reflection
band center wavelength $\lambda_{0}=\bar{n}p_{0}$ with the mean refractive
index $\bar{n}=(n_{o}+n_{e})/2$, $p_{0}=349$~nm.

\begin{figure}
\includegraphics[width=1\columnwidth]{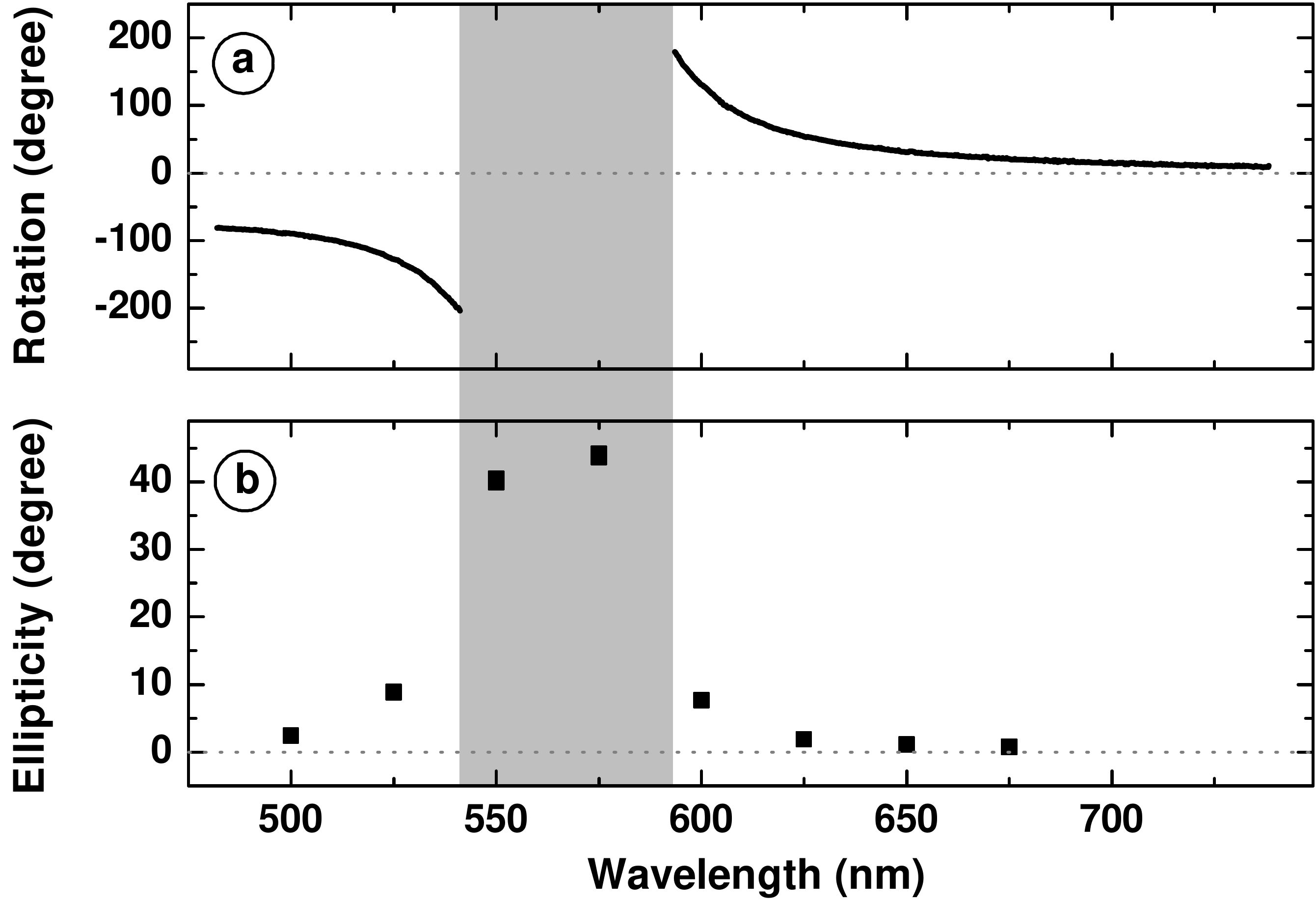}

\caption{\label{fig:rotcd}Measured optical rotation (a) and circular dichroism
(b) of the 16 $\mu$m thick cholesteric polymer. The optical rotation
shows a dispersive behavior, while the CD that of a resonance. The
CD is given as the ellipticity (circularly polarized light corresponds
to an ellipticity of 45 degrees). The gray box indicates the reflection
band from Fig.~\ref{fig:clc}. }
\end{figure}
We have characterized the optical activity of our cholesteric polymer
using standard diagnostics. The (spin-based) optical rotation and
circular dichroism of the sample are shown in Fig.~\ref{fig:rotcd}.
Conventional solutions of optically active molecules usually have
a specific optical rotation up to a few degrees/mm. Our cholesteric
polymer by far exceeds this value, close to the reflection band, the
optical rotation is around $2.3\times10^{4}$ degrees/mm. In this
band, the transmitted light is fully circularly polarized, and the
optical rotation cannot be determined. Therefore, we have measured
the circular dichroism, which can be done with high precision at any
spectral position. We give in Fig.~\ref{fig:rotcd}b the ellipticity
$\theta$ of the transmitted light, 45 degrees corresponds to fully
circularly polarized light.

\medskip{}

\begin{figure}
\includegraphics[width=1\columnwidth]{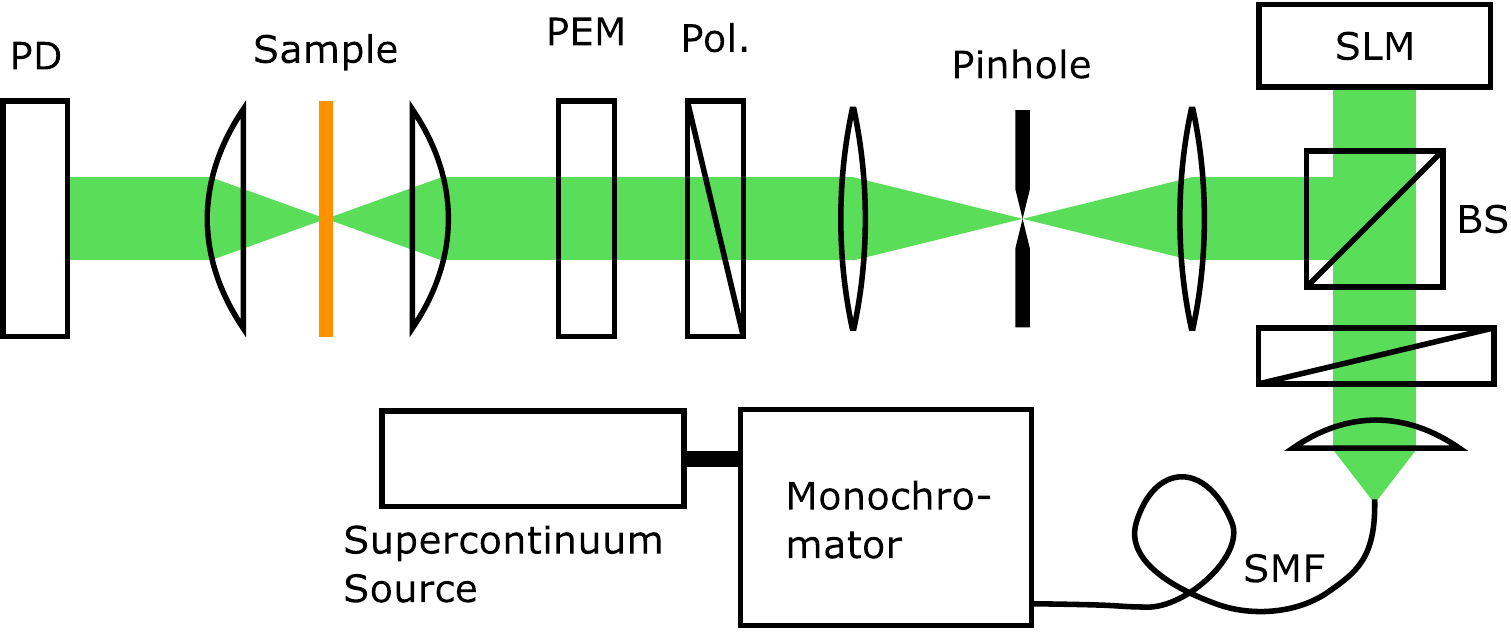}

\caption{\label{fig:exp}Scheme of the experiment. Pol.: linear polarizer;
PEM: photo-elastic modulator; SLM: spatial light modulator; PD: photo
diode; SMF: single-mode fiber. The PEM reference and the PD signal
are connected to a lock-in amplifier to determine the circular dichroism.}
\end{figure}

For our experiment to study the effect of OAM on the circular dichroism,
we use a tuneable light source a supercontinuum source (Fianium SC1060)
in combination with a monochromator (full width at half maximum FWHM
= 4~nm). After mode-filtering the output using a single-mode fiber,
we synthesize the OAM mode by holographic beam shaping with a phase-only
spatial light modulator (Fig.~\ref{fig:exp}) in conjunction with
a spatial filter (a pinhole in the Fourier plane of a lens). The polarization
of this beam is controlled with a Glan-Taylor polarizer and a photo-elastic
modulator (PEM), which is set to modulate between left- and right
circular polarization. The sample is at the mutual focus between two
microscopy objectives (Olympus MSPlan ULWD 50x, NA~=~0.55, the backside
aperture is filled by our light beam) forming a telescope. We record
the transmitted light with a photo diode (PD), this signal is fed
to a lock-in amplifier which is phase locked to the PEM polarization
modulation (50~kHz).

We test for a possible influence of OAM on the circular dichroism
by comparing the CD signal of light with OAM $\ell=+1$ or $\ell=-1$
at various spectral positions, within and outside of the cholesteric
reflection band. To isolate an OAM-induced effect, we introduce the
normalized differential circular dichroism ($CD=V_{AC}/V_{DC}$ is
the ratio of the measured polarization-differential transmitted intensity
normalized by the DC voltage):

\[
DCD=\frac{CD_{\ell=+1}-CD_{\ell=-1}}{CD_{\ell=+1}+CD_{\ell=-1}}
\]

\begin{figure}
\includegraphics[width=1\columnwidth]{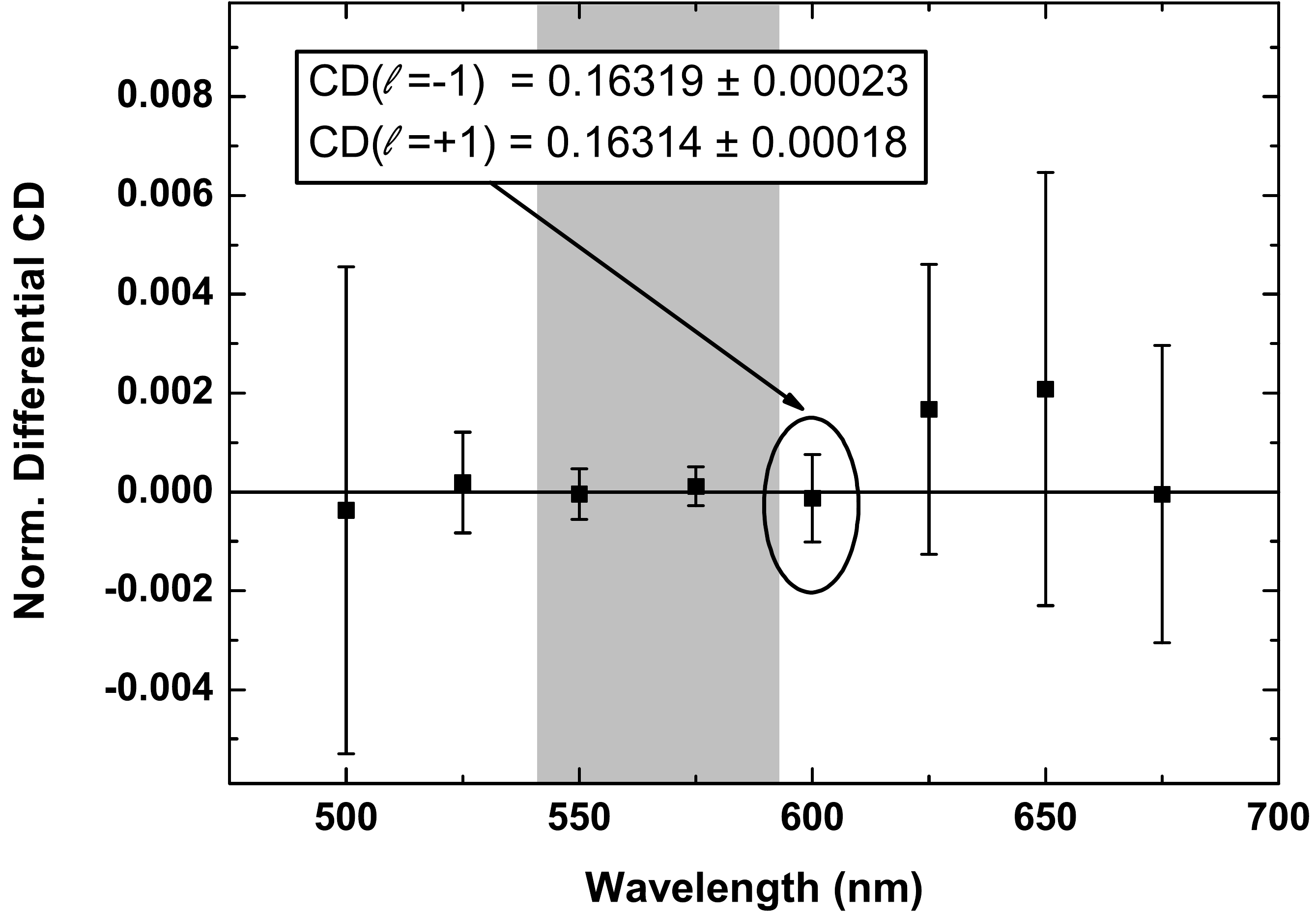}

\caption{\label{fig:oamcd}OAM-differential circular dichroism of the cholesteric
polymer. The measurement shows the normalized difference in circular
dichroism for $\ell=\text{{-}}1$ and $\ell=\text{{+}}1$ OAM beams.
The probe beam was focused using a 50x, NA~=~0.55 objective. A non-vanishing
differential CD would be an indication that OAM influences the circular
dichroism. The inset shows the measured CD signal at 600~nm wavelength
along with the uncertainty.}
\end{figure}

An influence of OAM on the CD would be demonstrated if $DCD\neq0$.
Fig.~\ref{fig:oamcd} shows the measured DCD. We conclude that OAM
has no effect here. For completeness we mention that the analogous
experiment with a collimated beam, not using the microscopic telescope,
shows equally a vanishing influence of OAM on the CD. We were able
to determine the circular dichroism with a relative uncertainty of
0.1\%, this is a significant improvement compared to the experiment
by Araoka et al \cite{araoka2005}, with an uncertainty of around
2\%. 

In conclusion, we did not observe interaction of light's orbital angular
momentum with a highly optically active cholesteric polymer in the
non-paraxial regime. This has two implications: Firstly, OAM light
does apparently not interact with the spatial resonances in a cholesteric
polymer. Secondly, this demonstrates that the conversion of spin angular
momentum to orbital angular momentum, as occurring at a high-NA lens,
can not be reversed: A lens can not simply convert light's OAM to
SAM. This agrees with recent theoretical results \cite{bliokh2010}.

\medskip{}

\begin{acknowledgments}
We acknowledge financial support by NWO and the EU STREP program 255914
(PHORBITECH).
\end{acknowledgments}
\bibliographystyle{apsrev4-1}
\bibliography{bibliography}

%merlin.mbs 2010-03-15 4.21a (PWD, AO, DPC)
\begin{thebibliography}{17}%
\makeatletter
\providecommand \@ifxundefined [1]{%
 \@ifx{#1\undefined}
}%
\providecommand \@ifnum [1]{%
 \ifnum #1\expandafter \@firstoftwo
 \else \expandafter \@secondoftwo
 \fi
}%
\providecommand \@ifx [1]{%
 \ifx #1\expandafter \@firstoftwo
 \else \expandafter \@secondoftwo
 \fi
}%
\providecommand \natexlab [1]{#1}%
\providecommand \enquote  [1]{``#1''}%
\providecommand \bibnamefont  [1]{#1}%
\providecommand \bibfnamefont [1]{#1}%
\providecommand \citenamefont [1]{#1}%
\providecommand \href@noop [0]{\@secondoftwo}%
\providecommand \href [0]{\begingroup \@sanitize@url \@href}%
\providecommand \@href[1]{\@@startlink{#1}\@@href}%
\providecommand \@@href[1]{\endgroup#1\@@endlink}%
\providecommand \@sanitize@url [0]{\catcode `\\12\catcode `\$12\catcode
  `\&12\catcode `\#12\catcode `\^12\catcode `\_12\catcode `\%12\relax}%
\providecommand \@@startlink[1]{}%
\providecommand \@@endlink[0]{}%
\providecommand \url  [0]{\begingroup\@sanitize@url \@url }%
\providecommand \@url [1]{\endgroup\@href {#1}{\urlprefix }}%
\providecommand \urlprefix  [0]{URL }%
\providecommand \Eprint [0]{\href }%
\@ifxundefined \urlstyle {%
  \providecommand \doi  [0]{\begingroup \@sanitize@url \@doi}%
  \providecommand \@doi [1]{\endgroup \@@startlink {\doibase
  #1}doi:\discretionary {}{}{}#1\@@endlink }%
}{%
  \providecommand \doi  [0]{doi:\discretionary{}{}{}\begingroup
  \urlstyle{rm}\Url }%
}%
\providecommand \doibase [0]{http://dx.doi.org/}%
\providecommand \Doi [0]{\begingroup \@sanitize@url \@Doi }%
\providecommand \@Doi  [1]{\endgroup\@@startlink{\doibase#1}\@@Doi}%
\providecommand \@@Doi [1]{#1\@@endlink}%
\providecommand \selectlanguage [0]{\@gobble}%
\providecommand \bibinfo  [0]{\@secondoftwo}%
\providecommand \bibfield  [0]{\@secondoftwo}%
\providecommand \translation [1]{[#1]}%
\providecommand \BibitemOpen [0]{}%
\providecommand \bibitemStop [0]{}%
\providecommand \bibitemNoStop [0]{.\EOS\space}%
\providecommand \EOS [0]{\spacefactor3000\relax}%
\providecommand \BibitemShut  [1]{\csname bibitem#1\endcsname}%
%</preamble>
\bibitem [{\citenamefont {Barron}(2004)}]{barron2004}%
  \BibitemOpen
  \bibfield  {author} {\bibinfo {author} {\bibfnamefont {L.~D.}\ \bibnamefont
  {Barron}},\ }\href@noop {} {\emph {\bibinfo {title} {{Molecular light
  scattering and optical activity}}}}\ (\bibinfo  {publisher} {Cambridge
  University Press},\ \bibinfo {year} {2004})\BibitemShut {NoStop}%
\bibitem [{\citenamefont {Allen}\ \emph {et~al.}(1992)\citenamefont {Allen},
  \citenamefont {Beijersbergen}, \citenamefont {Spreeuw},\ and\ \citenamefont
  {Woerdman}}]{allen1992}%
  \BibitemOpen
  \bibfield  {author} {\bibinfo {author} {\bibfnamefont {L.}~\bibnamefont
  {Allen}}, \bibinfo {author} {\bibfnamefont {M.~W.}\ \bibnamefont
  {Beijersbergen}}, \bibinfo {author} {\bibfnamefont {R.~J.~C.}\ \bibnamefont
  {Spreeuw}}, \ and\ \bibinfo {author} {\bibfnamefont {J.~P.}\ \bibnamefont
  {Woerdman}},\ }\Doi {10.1103/PhysRevA.45.8185} {\bibfield  {journal}
  {\bibinfo  {journal} {Phys. Rev. A},\ }\textbf {\bibinfo {volume} {45}},\
  \bibinfo {pages} {8185} (\bibinfo {year} {1992})}\BibitemShut {NoStop}%
\bibitem [{\citenamefont {Alexandrescu}\ \emph {et~al.}(2006)\citenamefont
  {Alexandrescu}, \citenamefont {Cojoc},\ and\ \citenamefont
  {Fabrizio}}]{alexandrescu2006}%
  \BibitemOpen
  \bibfield  {author} {\bibinfo {author} {\bibfnamefont {A.}~\bibnamefont
  {Alexandrescu}}, \bibinfo {author} {\bibfnamefont {D.}~\bibnamefont {Cojoc}},
  \ and\ \bibinfo {author} {\bibfnamefont {E.~D.}\ \bibnamefont {Fabrizio}},\
  }\Doi {10.1103/PhysRevLett.96.243001} {\bibfield  {journal} {\bibinfo
  {journal} {Phys. Rev. Lett.},\ }\textbf {\bibinfo {volume} {96}},\ \bibinfo
  {eid} {243001} (\bibinfo {year} {2006})}\BibitemShut {NoStop}%
\bibitem [{\citenamefont {J\'auregui}(2004)}]{jauregui2004}%
  \BibitemOpen
  \bibfield  {author} {\bibinfo {author} {\bibfnamefont {R.}~\bibnamefont
  {J\'auregui}},\ }\Doi {10.1103/PhysRevA.70.033415} {\bibfield  {journal}
  {\bibinfo  {journal} {Phys. Rev. A},\ }\textbf {\bibinfo {volume} {70}},\
  \bibinfo {pages} {033415} (\bibinfo {year} {2004})}\BibitemShut {NoStop}%
\bibitem [{\citenamefont {D\'avila~Romero}\ \emph {et~al.}(2002)\citenamefont
  {D\'avila~Romero}, \citenamefont {Andrews},\ and\ \citenamefont
  {Babiker}}]{romero2002}%
  \BibitemOpen
  \bibfield  {author} {\bibinfo {author} {\bibfnamefont {L.~C.}\ \bibnamefont
  {D\'avila~Romero}}, \bibinfo {author} {\bibfnamefont {D.~L.}\ \bibnamefont
  {Andrews}}, \ and\ \bibinfo {author} {\bibfnamefont {M.}~\bibnamefont
  {Babiker}},\ }\href@noop {} {\bibfield  {journal} {\bibinfo  {journal} {J.
  Opt. B},\ }\textbf {\bibinfo {volume} {4}},\ \bibinfo {pages} {S66} (\bibinfo
  {year} {2002})}\BibitemShut {NoStop}%
\bibitem [{\citenamefont {{Andrews}}\ \emph {et~al.}(2003)\citenamefont
  {{Andrews}}, \citenamefont {{D\'{a}vila~Romero}},\ and\ \citenamefont
  {{Babiker}}}]{andrews2003}%
  \BibitemOpen
  \bibfield  {author} {\bibinfo {author} {\bibfnamefont {D.~L.}\ \bibnamefont
  {{Andrews}}}, \bibinfo {author} {\bibfnamefont {L.~C.}\ \bibnamefont
  {{D\'{a}vila~Romero}}}, \ and\ \bibinfo {author} {\bibfnamefont
  {M.}~\bibnamefont {{Babiker}}},\ }\href@noop {} {\bibfield  {journal}
  {\bibinfo  {journal} {ArXiv Physics e-prints}} (\bibinfo {year} {2003})},\
  \Eprint {http://arxiv.org/abs/arXiv:physics/0305002} {arXiv:physics/0305002}
  \BibitemShut {NoStop}%
\bibitem [{\citenamefont {Andrews}\ \emph {et~al.}(2004)\citenamefont
  {Andrews}, \citenamefont {D\'{a}vila~Romero},\ and\ \citenamefont
  {Babiker}}]{andrews2004}%
  \BibitemOpen
  \bibfield  {author} {\bibinfo {author} {\bibfnamefont {D.~L.}\ \bibnamefont
  {Andrews}}, \bibinfo {author} {\bibfnamefont {L.~C.}\ \bibnamefont
  {D\'{a}vila~Romero}}, \ and\ \bibinfo {author} {\bibfnamefont
  {M.}~\bibnamefont {Babiker}},\ }\Doi {DOI: 10.1016/j.optcom.2004.03.093}
  {\bibfield  {journal} {\bibinfo  {journal} {Opt. Commun.},\ }\textbf
  {\bibinfo {volume} {237}},\ \bibinfo {pages} {133 } (\bibinfo {year}
  {2004})}\BibitemShut {NoStop}%
\bibitem [{\citenamefont {Babiker}\ \emph {et~al.}(2002)\citenamefont
  {Babiker}, \citenamefont {Bennett}, \citenamefont {Andrews},\ and\
  \citenamefont {D\'avila~Romero}}]{babiker2002}%
  \BibitemOpen
  \bibfield  {author} {\bibinfo {author} {\bibfnamefont {M.}~\bibnamefont
  {Babiker}}, \bibinfo {author} {\bibfnamefont {C.~R.}\ \bibnamefont
  {Bennett}}, \bibinfo {author} {\bibfnamefont {D.~L.}\ \bibnamefont
  {Andrews}}, \ and\ \bibinfo {author} {\bibfnamefont {L.~C.}\ \bibnamefont
  {D\'avila~Romero}},\ }\Doi {10.1103/PhysRevLett.89.143601} {\bibfield
  {journal} {\bibinfo  {journal} {Phys. Rev. Lett.},\ }\textbf {\bibinfo
  {volume} {89}},\ \bibinfo {pages} {143601} (\bibinfo {year}
  {2002})}\BibitemShut {NoStop}%
\bibitem [{\citenamefont {Araoka}\ \emph {et~al.}(2005)\citenamefont {Araoka},
  \citenamefont {Verbiest}, \citenamefont {Clays},\ and\ \citenamefont
  {Persoons}}]{araoka2005}%
  \BibitemOpen
  \bibfield  {author} {\bibinfo {author} {\bibfnamefont {F.}~\bibnamefont
  {Araoka}}, \bibinfo {author} {\bibfnamefont {T.}~\bibnamefont {Verbiest}},
  \bibinfo {author} {\bibfnamefont {K.}~\bibnamefont {Clays}}, \ and\ \bibinfo
  {author} {\bibfnamefont {A.}~\bibnamefont {Persoons}},\ }\Doi
  {10.1103/PhysRevA.71.055401} {\bibfield  {journal} {\bibinfo  {journal}
  {Phys. Rev. A},\ }\textbf {\bibinfo {volume} {71}},\ \bibinfo {pages}
  {055401} (\bibinfo {year} {2005})}\BibitemShut {NoStop}%
\bibitem [{\citenamefont {Barnett}\ and\ \citenamefont
  {Allen}(1994)}]{barnett1994}%
  \BibitemOpen
  \bibfield  {author} {\bibinfo {author} {\bibfnamefont {S.~M.}\ \bibnamefont
  {Barnett}}\ and\ \bibinfo {author} {\bibfnamefont {L.}~\bibnamefont
  {Allen}},\ }\Doi {DOI: 10.1016/0030-4018(94)90269-0} {\bibfield  {journal}
  {\bibinfo  {journal} {Opt. Commun.},\ }\textbf {\bibinfo {volume} {110}},\
  \bibinfo {pages} {670 } (\bibinfo {year} {1994})}\BibitemShut {NoStop}%
\bibitem [{\citenamefont {Bliokh}\ \emph {et~al.}(2010)\citenamefont {Bliokh},
  \citenamefont {Alonso}, \citenamefont {Ostrovskaya},\ and\ \citenamefont
  {Aiello}}]{bliokh2010}%
  \BibitemOpen
  \bibfield  {author} {\bibinfo {author} {\bibfnamefont {K.~Y.}\ \bibnamefont
  {Bliokh}}, \bibinfo {author} {\bibfnamefont {M.~A.}\ \bibnamefont {Alonso}},
  \bibinfo {author} {\bibfnamefont {E.~A.}\ \bibnamefont {Ostrovskaya}}, \ and\
  \bibinfo {author} {\bibfnamefont {A.}~\bibnamefont {Aiello}},\ }\Doi
  {10.1103/PhysRevA.82.063825} {\bibfield  {journal} {\bibinfo  {journal}
  {Phys. Rev. A},\ }\textbf {\bibinfo {volume} {82}},\ \bibinfo {pages}
  {063825} (\bibinfo {year} {2010})}\BibitemShut {NoStop}%
\bibitem [{\citenamefont {Nieminen}\ \emph {et~al.}(2008)\citenamefont
  {Nieminen}, \citenamefont {Stilgoe}, \citenamefont {Heckenberg},\ and\
  \citenamefont {Rubinsztein-Dunlop}}]{nieminen2008}%
  \BibitemOpen
  \bibfield  {author} {\bibinfo {author} {\bibfnamefont {T.~A.}\ \bibnamefont
  {Nieminen}}, \bibinfo {author} {\bibfnamefont {A.~B.}\ \bibnamefont
  {Stilgoe}}, \bibinfo {author} {\bibfnamefont {N.~R.}\ \bibnamefont
  {Heckenberg}}, \ and\ \bibinfo {author} {\bibfnamefont {H.}~\bibnamefont
  {Rubinsztein-Dunlop}},\ }\href@noop {} {\bibfield  {journal} {\bibinfo
  {journal} {J. Opt. A},\ }\textbf {\bibinfo {volume} {10}},\ \bibinfo {pages}
  {115005} (\bibinfo {year} {2008})}\BibitemShut {NoStop}%
\bibitem [{\citenamefont {Broer}\ and\ \citenamefont
  {Heynderickx}(1990)}]{broer1990}%
  \BibitemOpen
  \bibfield  {author} {\bibinfo {author} {\bibfnamefont {D.~J.}\ \bibnamefont
  {Broer}}\ and\ \bibinfo {author} {\bibfnamefont {I.}~\bibnamefont
  {Heynderickx}},\ }\href@noop {} {\bibfield  {journal} {\bibinfo  {journal}
  {Macromolecules},\ }\textbf {\bibinfo {volume} {23}},\ \bibinfo {pages}
  {2474} (\bibinfo {year} {1990})}\BibitemShut {NoStop}%
\bibitem [{\citenamefont {Hikmet}\ \emph {et~al.}(1991)\citenamefont {Hikmet},
  \citenamefont {Lub},\ and\ \citenamefont {Broer}}]{hikmet1991}%
  \BibitemOpen
  \bibfield  {author} {\bibinfo {author} {\bibfnamefont {R.~A.~M.}\
  \bibnamefont {Hikmet}}, \bibinfo {author} {\bibfnamefont {J.}~\bibnamefont
  {Lub}}, \ and\ \bibinfo {author} {\bibfnamefont {D.~J.}\ \bibnamefont
  {Broer}},\ }\href@noop {} {\bibfield  {journal} {\bibinfo  {journal} {Adv.
  Mater.},\ }\textbf {\bibinfo {volume} {3}},\ \bibinfo {pages} {392} (\bibinfo
  {year} {1991})}\BibitemShut {NoStop}%
\bibitem [{\citenamefont {Lub}\ \emph {et~al.}(1995)\citenamefont {Lub},
  \citenamefont {Broer}, \citenamefont {Hikmet},\ and\ \citenamefont
  {Nierop}}]{lub1995}%
  \BibitemOpen
  \bibfield  {author} {\bibinfo {author} {\bibfnamefont {J.}~\bibnamefont
  {Lub}}, \bibinfo {author} {\bibfnamefont {D.~J.}\ \bibnamefont {Broer}},
  \bibinfo {author} {\bibfnamefont {R.~A.~M.}\ \bibnamefont {Hikmet}}, \ and\
  \bibinfo {author} {\bibfnamefont {K.~G.~J.}\ \bibnamefont {Nierop}},\
  }\href@noop {} {\bibfield  {journal} {\bibinfo  {journal} {Liquid Crystals},\
  }\textbf {\bibinfo {volume} {18}},\ \bibinfo {pages} {319} (\bibinfo {year}
  {1995})}\BibitemShut {NoStop}%
\bibitem [{\citenamefont {St.~John}\ \emph {et~al.}(1995)\citenamefont
  {St.~John}, \citenamefont {Fritz}, \citenamefont {Lu},\ and\ \citenamefont
  {Yang}}]{stjohn1995}%
  \BibitemOpen
  \bibfield  {author} {\bibinfo {author} {\bibfnamefont {W.~D.}\ \bibnamefont
  {St.~John}}, \bibinfo {author} {\bibfnamefont {W.~J.}\ \bibnamefont {Fritz}},
  \bibinfo {author} {\bibfnamefont {Z.~J.}\ \bibnamefont {Lu}}, \ and\ \bibinfo
  {author} {\bibfnamefont {D.-K.}\ \bibnamefont {Yang}},\ }\Doi
  {10.1103/PhysRevE.51.1191} {\bibfield  {journal} {\bibinfo  {journal} {Phys.
  Rev. E},\ }\textbf {\bibinfo {volume} {51}},\ \bibinfo {pages} {1191}
  (\bibinfo {year} {1995})}\BibitemShut {NoStop}%
\bibitem [{\citenamefont {Berreman}\ and\ \citenamefont
  {Scheffer}(1970)}]{berreman1970}%
  \BibitemOpen
  \bibfield  {author} {\bibinfo {author} {\bibfnamefont {D.~W.}\ \bibnamefont
  {Berreman}}\ and\ \bibinfo {author} {\bibfnamefont {T.~J.}\ \bibnamefont
  {Scheffer}},\ }\Doi {10.1103/PhysRevLett.25.577} {\bibfield  {journal}
  {\bibinfo  {journal} {Phys. Rev. Lett.},\ }\textbf {\bibinfo {volume} {25}},\
  \bibinfo {pages} {577} (\bibinfo {year} {1970})}\BibitemShut {NoStop}%
\end{thebibliography}%

\newpage{}
\end{document}